\documentclass[sigconf,natbib=true]{acmart}
\usepackage{multirow}
\usepackage{makecell}
\usepackage{subfig}
\usepackage{bbding}
\usepackage{enumitem}
\usepackage{colortbl}
\usepackage{diagbox}

\setcopyright{acmcopyright}
\copyrightyear{2023}
\acmYear{2023}
\acmDOI{10.1145/xxxxxxx.xxxxxxx}

\acmConference[Woodstock '18]{Woodstock '18: ACM Symposium on Neural
  Gaze Detection}{June 03--05, 2018}{Woodstock, NY}
\acmBooktitle{Woodstock '18: ACM Symposium on Neural Gaze Detection,
  June 03--05, 2018, Woodstock, NY}
\acmPrice{15.00}
\acmISBN{978-1-4503-XXXX-X/18/06}

\begin{document}
\fancyhead{}
\title{A Comprehensive Summarization and Evaluation of Feature Refinement Modules for CTR Prediction}

\author{Fangye Wang}
\authornote{
Also with Shanghai Key Laboratory of Data Science, Shanghai Institute of Intelligent Electronics \& Systems, China.
}
\orcid{0000-0001-7216-1688}
\affiliation{
{
\normalsize
  \institution{School of Computer Science}
  \institution{Fudan University
  \city{Shanghai}
  \country{China}}
}
}
\email{fywang18@fudan.edu.cn}

\author{Hansu Gu}
{
\normalsize
\affiliation{
  \city{Seattle}
  \country{United States}}
}
\email{hansug@acm.org}

\author{Dongsheng Li}
\affiliation{
{
\normalsize
  \institution{Microsoft Research Asia}
  \city{Shanghai}
  \country{China}
 }
}
\email{dongsli@microsoft.com}

\author{Tun Lu}
\authornotemark[1]
\authornote{Corresponding author.}
\affiliation{
\normalsize
  \institution{School of Computer Science}
  \institution{Fudan University 
  \city{Shanghai} 
  \country{China}}
 }
\email{lutun@fudan.edu.cn}

\author{Peng Zhang}
\authornotemark[1]
\affiliation{
 \normalsize
  \institution{School of Computer Science}
  \institution{Fudan University
  \city{Shanghai}
  \country{China}
  }
 }
\email{zhangpeng_@fudan.edu.cn}

\author{Li Shang}
\authornotemark[1]
\affiliation{
{
  \normalsize
  \institution{School of Computer Science}
  \institution{Fudan University
  \city{Shanghai}
  \country{China}}
 }
}

\email{lishang@fudan.edu.cn}

\author{Ning Gu}
\authornotemark[1]
\affiliation{
{
  \normalsize
  \institution{School of Computer Science}
  \institution{Fudan University
  \city{Shanghai}
  \country{China}}
 }
}
\email{ninggu@fudan.edu.cn}

\renewcommand{\shortauthors}{Wang, et al.}

\begin{abstract}
Click-through rate (CTR) prediction is widely used in academia and industry. Most CTR tasks fall into a feature embedding \& feature interaction paradigm, where the accuracy of CTR prediction is mainly improved by designing practical feature interaction structures. However, recent studies have argued that the fixed feature embedding learned only through the embedding layer limits the performance of existing CTR models. Some works apply extra modules on top of the embedding layer to dynamically refine feature representations in different instances, making it effective and easy to integrate with existing CTR methods. Despite the promising results, there is a lack of a systematic review and summarization of this new promising direction on the CTR task. To fill this gap, we comprehensively summarize and define a new module, namely \textbf{feature refinement} (FR) module, that can be applied between feature embedding and interaction layers. We extract 14 FR modules from previous works, including instances where the FR module was proposed but not clearly defined or explained. We fully assess the effectiveness and compatibility of existing FR modules through comprehensive and extensive experiments with over 200 augmented models and over 4,000 runs for more than 15,000 GPU hours. The results offer insightful guidelines for researchers, and all benchmarking code and experimental results are open-sourced. In addition, we present a new architecture of assigning independent FR modules to separate sub-networks for parallel CTR models, as opposed to the conventional method of inserting a shared FR module on top of the embedding layer. Our approach is also supported by comprehensive experiments demonstrating its effectiveness.

\end{abstract}



\maketitle

\section{Introduction}
\label{sec:introduction}

\begin{figure}[t]
    \setlength{\abovecaptionskip}{0.2cm}
    \setlength{\belowcaptionskip}{-0.2cm}
    \centering
    \includegraphics[width=0.48\textwidth]{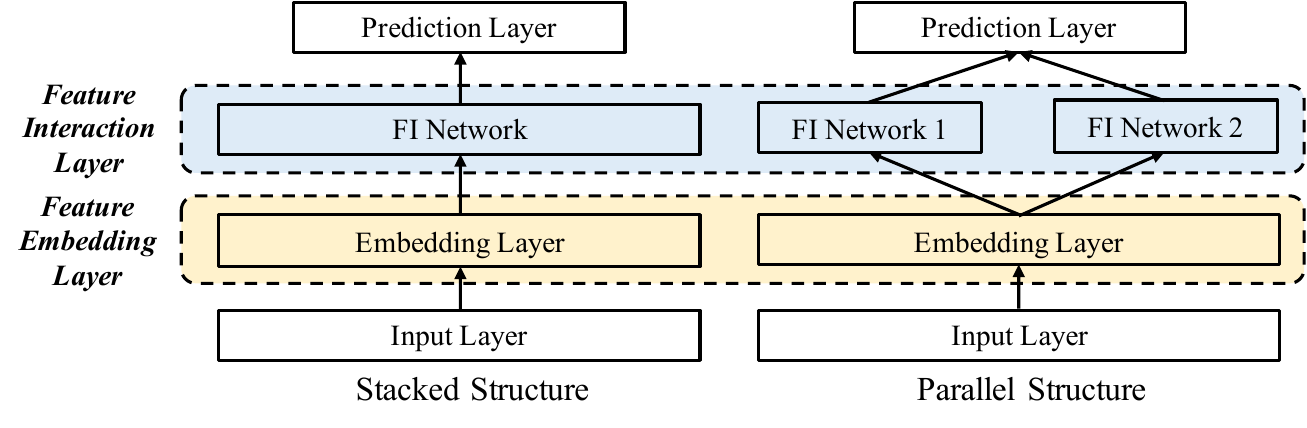}
    \caption{The two primary backbone structures of common CTR prediction models. \textit{Left}: stacked structure, e.g., NFM, PNN. \textit{Right}: parallel structure, e.g., DeepFM, DCN. Both of two structures contain FE layer, FI layer and prediction layer.} 
    \label{fig:stack_pa}
\end{figure}

Click-through rate (CTR) prediction is a crucial task in recommendation systems and online advertising ~\cite{ren2016user, chen2016deep}, which aims to predict the probability of a user clicking a recommended item or advertisement~\cite{graepel2010web, he2014practical}. The accuracy of CTR prediction impacts product ranking and ad placement, thus bringing enormous business value. As a result, there has been growing interest in improving CTR prediction accuracy in academia and industry. As summarized in ~\cite{zhang2021zs_deep, pan2021click_cl, luo2020network}, most CTR prediction models follow the same design paradigm: Feature Embedding (FE) layer, Feature Interaction (FI) layer, and Prediction layer. The FE layer converts the raw high-dimensional sparse features into a low-dimensional size~\cite{guo2021autodis, liu2019fgcnn, bian2022can}. The FI layer is placed on top of the FE layer to model the feature interactions and improve prediction performance effectively. Early research focuses on capturing low- or fixed-order feature interactions, e.g., FTRL~\cite{mcmahan2013ad}, FM~\cite{rendle2012factorization}, FFM~\cite{juan2016field}, and HOFM~\cite{blondel2016hofm}. With the advancement of deep neural network~\cite{wang2017deep_dcn,lecun2015deep, goodfellow2016deep}, deep learning based CTR models have been developed to capture informative arbitrary-order feature interactions and dramatically boost prediction accuracy. Well-designed FI structures, such as  DeepFM~\cite{guo2017deepfm}, xDeepFM~\cite{lian2018xdeepfm},  FiBiNet~\cite{huang2019fibinet}, AutoInt+~\cite{song2019autoint}, TFNet~\cite{wu2020tfnet}, DCNV2~\cite{wang2021dcnm}, have been proposed to learn explicit and implicit feature interactions jointly. As shown in Figure \ref{fig:stack_pa}, the above models follow two design patterns: stacked structure and parallel structure. Stacked structure models only deploy one FI network, while parallel structure models fuse two or more FI sub-networks for better performance.


Despite their success, the above CTR models have an inherent limitation where they can only learn fixed feature representations~\cite{wang2021contextnet, wu2020fafm, wang2022frnet}. This is because the model directly feed the embedding layer's output to the FI layer, so the same feature has identical representations across different input instances. For example, in the two instances, \{\textit{young, female, student, pink, skirt}\} and \{\textit{young, female, student, blue, notebook}\}, the feature "\textit{female}" has different impact on the click probability and should have different representations to reflect its specific contribution~\cite{yu2019input} more accurately. IFM~\cite{yu2019input} and DIFM~\cite{lu2021dual} improve FM by re-weighting basic features in different instances. Other studies~\cite{yu2019input, huang2020gatenet,wang2021contextnet, wang2022frnet} have also shown that instead of learning the fixed feature representations, using flexible feature representations can help improve the performance of CTR prediction models. For instance, FRNet~\cite{wang2022frnet} learns context-aware representations based on FM and achieves better performance than state-of-the-art (SOTA) models (xDeepFM~\cite{lian2018xdeepfm}, DCNV2~\cite{wang2021dcnm}) with fewer parameters and shorter training time. Additionally, a few works~\cite{huang2020gatenet, wang2021contextnet, wang2022frnet} propose different modules and successfully boost the performance of other SOTA CTR models~\cite{huang2019fibinet,guo2017deepfm, qu2018product}. These works share the common approach of learning dynamic feature representations by inserting feature refinement modules after the FE layer. These studies also illustrate that fixed feature representations indeed limit the model's performance. Unlike modeling feature interactions, learning dynamic feature representations is a fundamental and promising direction.

\begin{figure}[t]
    \setlength{\abovecaptionskip}{0.2cm}
    \setlength{\belowcaptionskip}{-0.2cm}
    \centering
    \includegraphics[width=0.48\textwidth]{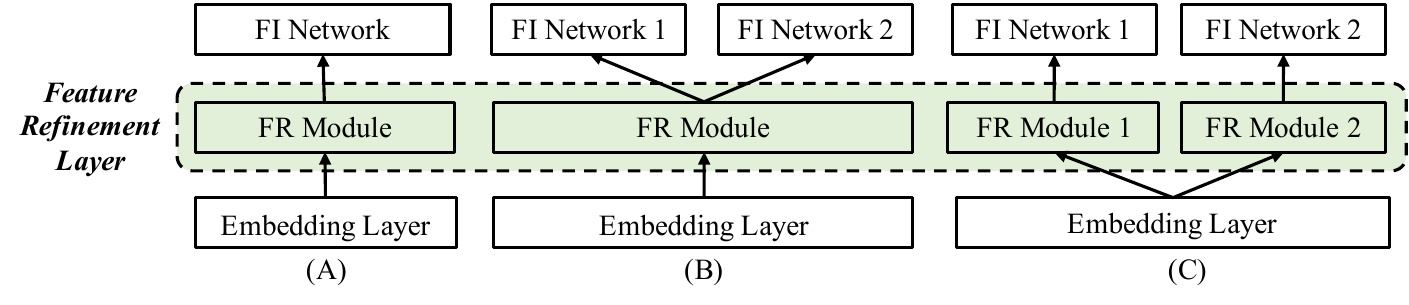}
    \caption{Two Integration patterns with FR module. (A) and (B): inserting a single FR module between FR and FI layer for stacked and parallel CTR models. (C): assigning two separate FR modules for different FI sub-network to generate discriminate feature distributions for parallel models.}
    \label{fig:add_refi}
\end{figure}

In this paper, we refer to these modules as \textbf{Feature Refinement (FR)} modules, which can dynamically refine the representations of each feature in different input instances based on their co-occurring features. Compared to creating new FI structures, the FR module can be more easily integrated into most CTR prediction models to improve their performance, as shown in Figure \ref{fig:add_refi}. However, comprehensive analysis and comparison of these FR modules are still lacking. We therefore identify and extract 14 FR modules from previous work, including instances where the FR module was proposed but not clearly defined or explained. Then we summarize 5 critical design properties of FR modules: information type, context-aware representation, weight, non-linearity and generation paradigm. To comprehensively compare and assess the effectiveness and compatibility of these FR modules, we integrate them into existing SOTA CTR prediction models and conduct a series of experiments. Specially, we create over 200 augmented models and run over 4,000 experiments, requiring more than 15,000 GPU hours in total for fair comparisons. We analyze the evaluation results from both qualitative and quantitative perspectives and suggest four promising research directions for future research. Our analysis and evaluation work can provide valuable guidance for researchers to advance the field of CTR prediction and beyond.

In summary, the contributions of this paper are as follows:

\begin{itemize}[leftmargin=0.4cm]
    \item This is the first survey that systematically reviews and summarizes FR modules. We identify 14 FR modules from existing CTR work and outline 5 essential properties that researchers should consider when they design FR modules.
    \item We fully assess the effectiveness and compatibility of existing FR modules through comprehensive and extensive experiments with over 200 augmented models and over 4,000 runs. The results offer insightful guidelines for researchers, and all benchmarking code and experimental results are open-sourced on GitHub at \url{https://github.com/codectr/RefineCTR}.
    \item We present a new architecture of having independent FR modules assigned to separate sub-networks, as opposed to the conventional method of inserting a shared feature refinement module on top of the embedding layer. Our approach is supported by comprehensive experiments demonstrating its effectiveness.
\end{itemize}

\section{Related Work}
\textbf{Feature Interaction.} 
Most FI-based CTR prdiction methods follow the Feature Embedding \& Feature Interaction paradigm. In recent years, modeling effective FI has become the commonly-adopted method to boost the performance of CTR prediction models. FM~\cite{rendle2012factorization} is one of the widely used models to capture pairwise feature second-order interactions via inner products. Due to its huge success, many FM-based models are proposed, e.g., NFM~\cite{he2017neural}, AFM~\cite{xiao2017attentional}, HFM~\cite{tay2019hfm}, RFM~\cite{punjabi2018rfm}, SEFM~\cite{lan2019sefm}, FMFM~\cite{sun2021fm2}, etc. Since the emergence of deep learning, many novel structures have been proposed to capture useful implicit or explicit interactions among features in different manners~\cite{zhu2020fuxictr}. Representative structures include inner product(e.g., FM~\cite{rendle2012factorization}, PNN~\cite{qu2018product}), bilinear interaction (e.g.,FiBiNET~\cite{huang2019fibinet}), attention mechanism (e.g., MIAN~\cite{song2019autoint}, DCAP~\cite{chen2021dcap}), CIN (e.g., xDeepFM~\cite{lian2018xdeepfm}), graph neural network (e.g., Fi-GNN~\cite{li2019fignn}, DG-ENN~\cite{guo2021dgenn}), convolution (e.g., CCPM~\cite{liu2015ccpm}, FGCNN~\cite{liu2019fgcnn}), field-aware interaction (e.g., FINT~\cite{zhao2021fint}), tensor-based interaction (e.g., TFNet~\cite{wu2020tfnet}), etc. According to the structure of CTR prediction models, we can divide them into stacked and parallel structures. Generally, stacked models only use one network or a stacked DNN to capture high-order interaction information~\cite{chen2021enhancing, song2019autoint}, e.g., NFM~\cite{he2017neural}, HoAFM~\cite{tao2020hoafm}, OENN~\cite{guo2019oenn}, AutoInt~\cite{song2019autoint}, etc. Meanwhile, parallel models leverage two or more FI sub-networks to capture explicit and implicit feature interactive
signals. Many parallel models integrate DNN to capture implicit interactive information, e.g., WDL~\cite{cheng2016wide}, DeepFM~\cite{guo2017deepfm}, AFN+~\cite{cheng2020adaptive} and DCNV2~\cite{wang2021dcnm}. Other models, like NON~\cite{luo2020network}, FED~\cite{zhao2020dimension}, JointCTR~\cite{yan2022jointctr}, fuse several interactive information for better performance. Generally, ensembled parallel models are superior to stacked ones.

\textbf{AutoML.} Designing effective structures for capturing useful FI information heavily relies on human experiences and expert knowledge. Hence, AutoML is adopted to construct proper feature interaction networks automatically. In the FE layer, AutoML can be used in the FE layer for searching embedding dimensions~\cite{zhao2021autodim, zhaok2021autoemb, liu2020automated} and embedding pruning~\cite{qu2022single, yan2021learning}. In the FI layer, AutoML is adopted to search useful feature interactions~\cite{liu2020autogroup, liu2020autofis,xie2021fives,khawar2020autofeature} and effective design pattern~\cite{meng2021autopi,wei2021autoias, zhao2021ameir}. In the prediction layer, AutoML can be used to search loss functions~\cite{zhao2021autoloss}. 

\textbf{Feature Refinement.} IFM~\cite{yu2019input} and DIFM~\cite{lu2021dual} improve FM by assigning dynamic feature importance to each feature based on each input instance. Other works propose different modules to refine feature representations and successfully improve the performance of existing CTR models significantly, e.g., ContextNet~\cite{wang2021contextnet}, GateNet~\cite{huang2020gatenet}, FRNet~\cite{wang2022frnet}. These works propose a new direction to boost the performance of CTR prediction models, i.e., refining feature representations. However, most modules are only proposed to improve individual models, and those works evaluate their proposed modules based on different base models or datasets. Therefore, a complete and fair evaluation of those modules is lacking. Our work fills this gap by providing a comprehensive summarization of this novel and promising direction for CTR prediction. 

\textbf{Existing CTR Prediction Surveys and Benchmarks.} Due to the broad application of CTR prediction models in various fields and the rapid emergence of new CTR prediction models, several  surveys have been conducted to summarize existing CTR models. \citet{zhang2021zs_deep} provides a review of the development of deep learning models for CTR prediction tasks, which elaborates on several primary research directions of the current CTR prediction tasks, including designing useful feature interactions, modeling user behavior sequences and automated architecture search. \cite{wang2020zs_survey}, and \cite{yang2022zs_click} make a systematic literature review on SOTA and the latest CTR prediction models, with a special focus on modeling complex and useful feature interactions. Unfortunately, there is still a lack of standardized benchmarks and inconsistent evaluation criteria for CTR prediction research~\cite {zhu2022bars,zhu2020fuxictr}. The summary of the different model performances in ~\cite{yang2022zs_click} also confirms this issue. BARS~\cite{zhu2022bars} and FuxiCTR~\cite{zhu2020fuxictr} standardize the evaluation protocols and provide the most comprehensive benchmarking results that rigorously compare existing models. Additionally, ~\cite{zheng2022automl} summarizes existing CTR models which utilize AutoML to automatically search for the proper candidates for different parts in the CTR task.

Above studies mainly summarized and compared the performance of single CTR models, but our work first summarize and evaluate the effectiveness of FR modules on the CTR prediction task. Given the rising popularity and potential of CTR prediction and the steady flow of novel research contributions (i.e., FI structures) in this area, a comprehensive evaluation of FR modules for improving CTR prediction will be of high scientific and practical value. Our work comprehensively analyzes the current research on FR modules to bridge the gap between the FE and FI layers. Therefore, for researchers or companies that are applying CTR models, it provides a convenient and effective way to improve the performance of existing models. 

\section{PRELIMINARIES}
The problem of CTR prediction is a binary classification task that aims to predict the probability that a user will click a recommended item based on the user and item features. The input data for CTR task are usually large-scale and highly sparse~\cite{liu2019fgcnn, qu2018product, rendle2012factorization}, and are represented in a multi-field categorical form~\cite{pan2021click_cl, huang2019fibinet, liu2019fgcnn}. Suppose there are $F$ different fields and $M$ features totally, where each field may contain multiple features, and each feature belongs to only one field~\cite{wang2022frnet, luo2020network}. Each input instance is formally represented by a high-dimensional sparse (binary) vector via one-hot encoding~\cite{liu2020autogroup, liu2019fgcnn}. Each instance for CTR prediction can be represented by $\left\{ \mathbf{x}_i, y_i \right\}$, where $\mathbf{x}_i$ is a sparse high-dimensional vector represented by one-hot encoding and $y_i\in{\{0,1\}}$ (click or not) is the true label. For example, an instance can be represented by:
\begin{align}
\underset{Field\ 1=Computer\,\,}{\underbrace{\left[ 0,...,1,0 \right] }}
\underset{Field\ 2=White\,\,}{\underbrace{\left[ 1,...,0 \right] }}
\underset{Field\ 3=18\,\,}{\underbrace{\left[ 0,...,1 \right] }}
\mathbf{...}
\underset{Field\ F=Student}{\underbrace{\left[ 1,0,...,0 \right]. }}
\end{align}

Most deep learning-based CTR prediction models have three fundamental layers from bottom to top: Feature Embedding layer, Feature Interaction layer, and Prediction layer, as shown in Figure \ref{fig:stack_pa}. We formally introduce the typical three layers in CTR prediction models as follows:

\textbf{{Feature Embedding layer.}} 
Since the feature representations of the categorical features are very sparse and high-dimensional, an embedding layer is employed to transform them into a dense low-dimensional embedding matrix $\mathbf{E}=[\mathbf{e}_1;\mathbf{e}_2;...;\mathbf{e}_f] \in \mathbb{R}^{F \times D}$, where $D$ is the dimension size, $\mathbf{e}_i$ is the representation vector of the i-th feature.

\textbf{{Feature Interaction layer.}} 
\label{sec:fi_cncoder}
The feature interaction layer aims to model effective feature interactions information. Formally, it transforms the embedding matrix $\mathbf{E}$ to a compact interaction vector $\mathbf{\mathit{h}}_i$, generated by $\mathbf{\mathit{h}}_i = FI(\mathbf{E})$. $FI(\mathbf{E})$ is the core FI structure. In the CTR prediction task, it is an essential component to boost prediction performance. For parallel models, $h_i$ integrates several sources of FI information.  

\textbf{{Prediction layer.}} The prediction layer outputs the final prediction probability $\hat{y_i} = \sigma(f(h_i)) \in [0, 1]$ based on the compact representations $\mathbf{\mathit{h}}_i$ from the FI layer. Usually, the function $f(\cdot)$ is a logistic regression module or DNN networks, and a Sigmoid function $\sigma(x)=1 /(1+\exp (-x))$ map the output to $[0, 1]$. 

Finally, the commonly adopted loss function is defined as the following:
\begin{align}
\textstyle
\mathcal{L}_{ctr} =-\frac{1}{N}\sum_{i=1}^N 
\left({y_i} 
    \log\left(\hat{y}_i\right) 
    +\left( 1-y_i \right) \log \left( 1- \hat{y}_i \right)
\right),
\end{align}
 where $y_i$ is the ground truth label, and $N$ is the number of training samples.

\section{Feature Refinement} 
\subsection{Feature Refinement}

\textbf{DEFINITION}
Feature Refinement is a feature representation learning methods, which can dynamically adjust the feature representations based on their co-occurring feature information across different input instances. Formally, given the raw feature embeddings $\mathbf{E} \in \mathbb{R}^{F \times D}$, the refined feature representations can be calculated as the following: 
\begin{align}
    \mathbf{E}_{fr} = FR(\mathbf{E}) \in \mathbb{R}^{F \times D},
\end{align}
where $FR(\cdot)$ is a FR module that generates the refined feature representations $\mathbf{E}_{fr}$. As the refined representations have the same dimensions as the raw representations, FR modules can be integrated into most existing CTR prediction models in a plug-and-play fashion. Additionally, for a specific feature (e.g., \textit{female}), FR module can generate dynamical refined representations in different instances which contains the feature.



\subsection{Representative Feature Refinement Modules}
We briefly introduce 14 FR modules extracted from existing works.

\textbf{Factor Estimating Net (FEN)}. In IFM~\cite{yu2019input}, FEN is the first work to explicitly consider the different importance of each feature across different input instances, which utilizes DNN and Softmax to learn vector-level weights dynamically. 

\textbf{Squeeze-Excitation Network (SENET)}. SENET is mainly used in image classification task~\cite{hu2018squeeze, roy2018senet_recalibrating}. FiBiNET~\cite{huang2019fibinet} leverages it to learn the importance of vector-wise feature by performing Squeeze, Excitation, and Re-Weight steps over the original representations.

\textbf{Field-wise network (FWN)}. In NON~\cite{luo2020network}, FWN is proposed to adequately capture the characteristics within each field, which is called intra-field information.

\textbf{Dimension Relation Module (DRM)}. FED~\cite{zhao2020dimension} designs DRM to learn the feature interactions through operations on all embedding features without considering the relationships between the learned latent properties. 

\textbf{Dual Factor Estimating Net (DFEN)}. DIFM~\cite{lu2021dual} designs DFEN based on FEN, which improves FM by re-weighting feature representations. Different from FEN, DFEN employs multi-head attention and ResNet~\cite{he2016resnet} mechanisms to learn vector-level weights.

\textbf{Feature Adjustment Layer (FAL)}. Inspired by IFM, FaFM~\cite{wu2020fafm} designs FAL to improve FM, which naturally integrates vector-level learning and bit-level learning modules. Notably, FAL has a user and item interaction-aware structure, which requires that the input data not anonymous. Therefore, we do not evaluate FAL in the experiments as Criteo dataset is anonymous. 

\textbf{VGate and BGate}. GateNet ~\cite{huang2020gatenet} constructs vector- or bit-wise gates (i.e., VGate or BGate) and provides a learnable gating module to selects salient information from the feature or element level.

\textbf{Self-Attention (SelfAtt)}. InterHAt~\cite{li2020interpretable} considers the relationship among different features and utilizes multi-head self-attention mechanism to generate feature representations dynamically. 

\textbf{TCE and PFFN}. In ContextNet~\cite{wang2021contextnet}, TCE can dynamically generate feature representations according to other contextual features in the same instance. Based on TCE, PFFN stacks DNN and layer normalization to refine high-order feature representations.

\textbf{Gated Feature Refinement Layer (GFRL)}. In MCRF~\cite{wang2022mcrf}, GFRL generates an additional set of auxiliary feature representations and selects important information from the original features. In addition, GFRL integrates intra-field information with contextual information to learn bit-level information by Sigmoid activation.

\textbf{FRNet-V and FRNet-B}. FRNet~\cite{wang2022frnet} fuses contextual information with self-attention mechanism and adds complementary feature representations to the original representations to alleviate the linear relationship problem in other FR models (e.g., FEN, DFEN). FRNet provides two variants, FRNet-V and FRNet-B, to learn vector-level weights and bit-level weights, respectively.


\begin{table*}
    \setlength{\abovecaptionskip}{0.2cm}
    \setlength{\belowcaptionskip}{-0.2cm}
\centering
\caption{Summary of five key properties of feature refinement modules.}
\label{tab:all_feature}
\scalebox{0.80}{
\begin{tabular}{ccc|cccc|c|ccc|c|c} 
\hline
\hline
\multirow{2}{*}{Year} & \multirow{2}{*}{Module}& \multirow{2}{*}{Literature} & \multicolumn{4}{c|}{Information Type} & \multirow{2}{*}{\makecell[c]{Context-Aware}}   & \multicolumn{3}{c|}{Weight}  & \multirow{2}{*}{Non-linearity} & \multirow{2}{*}{\makecell[c]{Generation\\Paradigm}} \\ 
\cline{4-7} \cline{9-11}
&   & &IF & CF& CI& OI& & Granularity & Activation & Range&&\\ 
\hline
2019                  & FEN &IFM~\cite{yu2019input} &&& $\surd$ && $\surd$                   & Vector      & Softmax    & [0, 1] & $\times$                   & Selection\\
2019                  & SENET &FiBiNet~\cite{huang2019fibinet} &&&& $\surd$  & $\surd$  & Vector      & ReLU       & [0,$\infty$)  & $\times$& Selection \\
2020                  & FWN&NON~\cite{luo2020network}        & $\surd$  &&&     & $\times$            & Bit         & ReLU       & [0,$\infty$) & $\times$&Selection    \\
2020                  & DFEN&DIFM~\cite{lu2021dual}   && $\surd$ & $\surd$ &     & $\surd$               & Vector      & ReLU       & [0,$\infty$)   & $\times$& Selection \\
2020                  & DRM&FED~\cite{zhao2020dimension}   &&&& $\surd$& $\surd$                 & -           & -          & - & $\surd$& Transformation  \\
2020                  & FAL&FaFM~\cite{wu2020fafm}   && $\surd$ & $\surd$ &     & $\surd$                 & Bit         & ReLU       & [0,$\infty$)& $\surd$& Selection    \\
2020                  & VGate&GateNet~\cite{huang2020gatenet}    & $\surd$  &&& & $\times$              & Vector      & Identity     & (-$\infty$,$\infty$) & $\times$& Selection\\
2020                  & BGate&GateNet~\cite{huang2020gatenet}   & $\surd$  &&&& $\times$               & Bit         & Identity     & (-$\infty$,$\infty$) & $\times$& Selection\\
2020                  & SelfAtt&InterHAt~\cite{li2020interpretable}  && $\surd$ &&& $\surd$              & -           & -& -    & $\surd$& Transformation\\
2021                  & TCE&ContextNet~\cite{wang2021contextnet}   &&& $\surd$ & & $\surd$                 & Bit         & Identity     & (-$\infty$,$\infty$)& $\surd$& Selection \\
2021                  & PFFN&ContextNet~\cite{wang2021contextnet}  & $\surd$  && $\surd$ && $\surd$                 & -           & -& -    & $\surd$& Transformation\\
2022                  & GFRL&MCRF~\cite{wang2022mcrf} & $\surd$  && $\surd$ && $\surd$                  & Bit         & Sigmoid    & [0,1]  & $\surd$& Composite\\
2022                  & FRNet-V&FRNet~\cite{wang2022frnet}    && $\surd$ & $\surd$ && $\surd$             & Vector      & Sigmoid    & [0,1]  & $\surd$& Composite \\
2022                  & FRNet-B&FRNet~\cite{wang2022frnet}   && $\surd$ & $\surd$ && $\surd$             & Bit         & Sigmoid    & [0,1]  & $\surd$& Composite\\
\hline
\hline
\end{tabular}
}
\end{table*}

\subsection{Five Key Properties}
FR modules aim to refine features through three main processes: 1) improving the quality of raw feature embeddings, 2) designing weight learning functions to adjust the importance of each feature embedding, and 3) combining raw embeddings and learned weights to generate final refined representations. Table \ref{tab:all_feature} summarizes the five properties of FR modules.

The first two properties \textit{Information Type} and \textit{Context-Aware Representation} focus on improving the feature embedding quality. 

\textbf{Information Type:} 
Four types of information are used in FR modules to improve feature embeddings. 

1) \textit{Intra-field information (IF)}. Features within a field contain relevant information as they belong to the same field. However, there are still differences between these features that need to be captured. For instance, while `advertiser\_id' and `user\_id' are both parts of the "ID" field, they have distinct identities as noted in~\cite{luo2020network}. The information, whether a specific ID is an advertiser or a user, may help the following FI layer improve prediction accuracy. Each field is assigned with a $DNN_{i}$ to capture IF:
\begin{align}
    \mathbf{e}^{\prime}_i = DNN_i(\mathbf{e}_i) \in \mathbb{R}^{D}.
    \label{equ: intra_field info}
\end{align}
Since each instance has $F$ fields, we initialize F parallel DNNs. Practically, these DNNs can be computed in parallel to speedup~\cite{zhao2021non, wang2022mcrf}.

2) \textit{Cross-feature information (CF).} CF deals with the relationships between different features. Each feature is influenced by its co-occurring features, so a few methods refine its representations by considering its co-occurring features. Generally, the multi-head self-attention mechanism~\cite{vaswani2017attention} is a widely-used method, which refines feature representation by explicitly integrating the inter-dependencies among features (e.g., InterHAt~\cite{li2020interpretable}, FRNet~\cite{wang2022frnet}). The above process is represented by:
\begin{gather}
   \mathbf{E}_{att} = \operatorname{ReLU}(\operatorname{FeedForward}([\mathbf{H}_1;\mathbf{H}_2;...;\mathbf{H}_h]\mathbf{W}^{O})), \\
   \mathbf{H}_i=\operatorname{SoftMax}_i\left(
   \mathbf{Q_iK_i}^T/\sqrt{d_K}
   \right) \mathbf{V}_i, \\
   \mathbf{Q_i}= \mathbf{E}\mathbf{W}^Q_i, \mathbf{K_i}= \mathbf{E}\mathbf{W}^K_i, \mathbf{V_i}=\mathbf{E}\mathbf{W}^V_i,
\end{gather}
where the parameter matrices $\mathbf{W}^Q_i$, $\mathbf{W}^K_i$, $\mathbf{W}^V_i\in\mathbb{R}^{D \times d_K}$ and $\mathbf{W}^O \in \mathbb{R}^{hd_K \times D}$, $\mathbf{H}_i\in\mathbb{R}^{F \times d_K}$, $h$ is the number of attention heads. 

3) \textit{Contextual information (CI):} 
Contextual information is unique to each input instance and it is determined by all features combined. Typically CI is represented as a $D$ dimension vector generated by condensing all features information. Formally, 
\begin{align}
\mathbf{V}_{c} = F_{con}(\mathbf{E}) \in \mathbb{R}^{D}, 
\end{align}
where $F_{con}(\cdot)$ is the function that condenses all features $\mathbf{E}$ into CI, which can be implemented using DNN. The CI of the two instances can be significantly different, even if only one feature differs. Unlike CF, CI can not directly adjust the representation of a specific feature. A few works incorporate CI to IF or CF, allowing each feature to further identify the important cross-instance CI and learn more distinguishable representations based on varying contexts.

4) \textit{Other information (OI)}: Different from the above information, we also list several less commonly used information types. DRM considers all the relations among the learned latent space, while SENET focuses on the summary statistics of each field embedding, i.e., max or mean information of the original embedding. 

\textbf{Context-Aware Representation:} 
In addition to information type, the context-awareness of the embedding is also a design factor. To allow FR modules to refine embeddings dynamically, the same feature should have different representations across different instances. FWN, VGate and BGate are unable to generate context-aware representations since they only capture Intra-field information. As stated in Eq.\ref{equ: intra_field info}, since the feature embedding $\mathbf{e}_i$ of one feature and the parameters in $DNN_i$ are constant, the refined feature representations for a specific feature in different instances are still fixed. Therefore, generating context-aware representations requires a connection with other co-occurring features, e.g., integrating cross-feature or contextual information. A Visualization of several typical context-aware representations is shown in Figure \ref{fig:vis}. 

The next two properties \textit{Weight} and \textit{Non-linearity} focus on designing effective weighting strategy to adjust the importance of feature embeddings.

\begin{figure}[t]
    \setlength{\abovecaptionskip}{0.2cm}
    \setlength{\belowcaptionskip}{-0.2cm}
    \centering
    \includegraphics[width=0.38\textwidth]{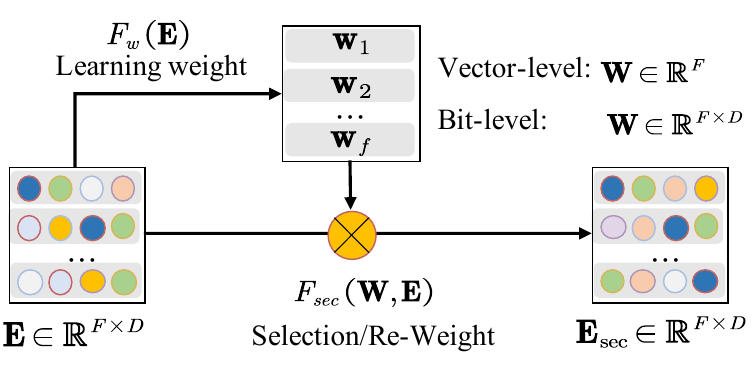}
    \caption{The process of adjusting feature representations by learning weight.}
    \label{fig:reweight}
\end{figure}

\textbf{Weight:} 
The performance of CTR prediction models is limited by the fixed feature representations, as the same feature is usually not equally useful in different input instances. Hence, most works learn and assign different weight to each feature in different instances to adjust feature representations dynamically, as shown in Figure~\ref{fig:reweight}. Based on the granularity of the weights, there are vector-level and bit-level weights, where vector-level assign a weight scalar to each feature vector and bit-level learn a weight for each bit of feature vector. We represent them as follows:
\begin{gather}
\mathbf{E}_{sec} = F_{sec}(\mathbf{W},\mathbf{E}) = [\mathbf{w}_1 \cdot \mathbf{e}_1, \mathbf{w}_2 \cdot \mathbf{e}_2, ..., \mathbf{w}_f \cdot \mathbf{e}_f], \\
\mathbf{W} = F_{w}(\mathbf{E})=[\mathbf{w}_1,\mathbf{w}_2,..., \mathbf{w}_f],
\label{equ:gate}
\end{gather}
where $F_{w}(\mathbf{E})$ and $F_{sec}(\mathbf{W},\mathbf{E})$ are the weight learning and feature selection function. $\mathbf{W}$ is the learned weight matrix, $\mathbf{E}$ and $\mathbf{E}_{sec}$ are the raw and refined feature representations. Specifically, for vector-level weights such as FEN and DFEN, $\mathbf{w}_i \in \mathbb{R}^{1}$, $\mathbf{W} \in \mathbb{R}^{F}$. For bit-level weight such as TCE and BGate, $\mathbf{w}_i \in \mathbb{R}^{D}$, $\mathbf{W} \in \mathbb{R}^{F \times D}$. The vector-level weights can be viewed as assigning the same weight to each bit of the feature representation. Compared with vector-level weight, bit-level weight is more fine-grained and generally works better. However, the vector-level weights provide better interpretability since the learned weights reflect the importance of the corresponding features. 

Although the weight learning functions $F_{w}(\mathbf{E})$ of various modules are significantly different, they all utilize an activation function at the final step, which impacts the weight range. Commonly used activation functions include: Sigmoid, Softmax, ReLU and Linear activations. Sigmoid and Softmax map the feature weights into [0, 1]. ReLU maps weights to [0, $\infty$), and linear activation function map weights to (-$\infty$, $\infty$). Most works~\cite{huang2020gatenet, lu2021dual} choose the most suitable activation function through empirical experiments.

\textbf{Non-linearity:} 
It indicates whether there are linear relationships between all corresponding refined representations of a specific feature and its original embedding. Given an embedding $e_i$ for a specific feature (e.g., \textit{female}), the vector-level weight is a scalar $m$ that applies to the $D$ dimensions of $\mathbf{e}_i$. For different instances containing the given feature, their corresponding refined representations are generated based on the raw vector $\mathbf{e}_i$ by assigning varying vector-level weights. Hence, all the refined $D$-dimension representation is linear to the original embedding $\mathbf{e}_i$, limiting the flexibility of the refined representations and decreasing their performance of them, e.g., FEN and SENET. Conversely, bit-level weights can generate non-linear feature representations and get better performance, which assigns different values to each bit of feature embedding $\mathbf{e}_i$. Figure \ref{fig:vis} visualizes the graph of linearity or non-linearity.


\textbf{Generation Paradigm:} 
The final property in FR design is the combination of improved features embeddings and weights to generate final refined feature representations $\mathbf{E}_{fr}$. There are three main paradigms: Selection, Transformation, and Composite.
\textbf{Selection} involves applying the weight matrix to the embeddings $\mathbf{E}$ to select the more important information, as depicted in Figure \ref{fig:reweight}.
\textbf{Transformation} means the refined representations are directly transformed from the embeddings, e.g., SelfAtt. without applying weights to the embedding to generate refined representations. 
\textbf{Composite} involves both selecting important information using the weight matrix and generating a group of complementary representations to compensate for the unselected information, as in GFRL, FRNet-B. Simplified, composite can be shown as follows:
\begin{gather}
    \mathbf{E}_{fr} = \mathbf{E} \cdot \mathbf{W} + \mathbf{E}_{com} \cdot (1-\mathbf{W}),\\
    \mathbf{E}_{com} = F_{com}(\mathbf{E}) \in \mathbb{R}^{F\times D},
 \end{gather}   
where $\mathbf{E}_{com}$ is the complementary representations and $F_{com}(\cdot)$ is the function to generate $\mathbf{E}_{com}$. Usually, the weight matrix $\mathbf{W}$ is generated with the Sigmoid function with range [0,1]. Intuitively, weight matrix 
$\mathbf{W}$ performs as a soft gate to select information from raw or complementary representations.

\subsection{New Integration Pattern}
Most CTR models~\cite{guo2017deepfm, wang2017deep_dcn, lian2018xdeepfm} share the same embedding layer for different sub-networks in both stacked and parallel models. The dimension of output embedding generated by FR module is the same as raw embedding, making it compatible with existing CTR models and can be easily integrated to improve model's performance. Most recent research~\cite{yu2019input, huang2020gatenet, wang2022frnet} also uses a shared FR module between FE layer and FI layer for stacked models in Figure~\ref{fig:add_refi}(A), and for parallel models shown in Figure~\ref{fig:add_refi}(B). However, different FI sub-networks require discriminative feature embedding distributions to capture different feature interactions information more targeted, which is ignored by existing parallel models. Although initializing a new set of feature embedding is the most widely used method to solve this issue, it results in huge memory consumption, e.g., FFM~\cite{juan2016field}, ONN~\cite{yang2020operation}, FAT-DeepFFM~\cite{zhang2019fat}. 

We present a novel design pattern to integrate FR modules with existing CTR prediction models that minimizes memory consumption. This is achieved by assigning separate FR modules for each FI sub-networks in parallel models, as shown in Figure \ref{fig:add_refi}(C). Compared with initializing a new set of feature embedding, the memory consumption of our proposed approach is minimal. We evaluate the effectiveness of this design pattern in Section~\ref{exp:newdesign}.

\section{EXPERIMENTAL Analysis}
\label{sec:experiment}
\subsection{Experiment Setup}
\textbf{Datasets:} 
We use two widely used and public datasets. (1) \textbf{Criteo}~\cite{criteo} is a popular and industrial dataset for CTR prediction, which comprises 45 million users’ clicking records on displayed ads. It contains 26 categorical feature fields and 13 numerical feature fields. (2) \textbf{Frappe}~\cite{frappe} contains mobile app usage logs from users in different contexts. The target value indicates whether the user has used the app or not. The datasets statistics are summarized in Table \ref{Tab.dataset}. 

\textbf{Data Preparation:}
First, we remove the infrequent feature (occurring less than threshold instances) and treat them as a single feature “<unknown>”, where the threshold is set to 10 for both Criteo dataset. Second, we normalize numerical values by the function $discrete(x) = \lfloor log^2{(x)} \rfloor$, where $\lfloor \textbf{.} \rfloor$ is the floor function, which is proposed by the winner of Criteo~\cite{winner}. Third, we randomly split Criteo and Frappe into 8:1:1 as the training, validation, and testing datasets, and control the random seed (i.e., seed=2022) for splitting. 

\textbf{Evaluation Metrics:}
We employ two commonly-used metrics, AUC and Logloss, to evaluate the performance of each CTR prediction model. \textbf{AUC} is the Area Under the ROC Curve. \textbf{Logloss} is the binary cross-entropy loss. Notably, an improvement of AUC at \textbf{0.001-level} is generally considered practically significant for an industrial CTR prediction task ~\cite{chen2021enhancing, guo2017deepfm, khawar2020autofeature, wang2021dcnm, liu2019fgcnn}. 

\textbf{Base and Augmented Models:}
We choose several representative and widely-cited models as base models to evaluate the performance of different FR modules. We first include representative stacked models: FM~\cite{rendle2012factorization}, CrossNet(CN)~\cite{wang2017deep_dcn}, CrossNetV2 (CN2)~\cite{wang2021dcnm}, and AFN ~\cite{cheng2020adaptive}, where these CTR prediction models can capture low- and high-order feature interactions explicitly or implicitly. In addition, four representative parallel models, including DeepFM~\cite{guo2017deepfm}, DCN~\cite{wang2017deep_dcn}, DCNV2~\cite{wang2021dcnm}, and AFN+~\cite{cheng2020adaptive}, can fuse two kinds of FI information by integrating DNN sub-network. Previous surveys~\cite{zhu2022bars, zhu2020fuxictr} have verified that the above CTR prediction models have achieved SOTA performance. Meanwhile, we also test other CTR prediction models, e.g., NFM~\cite{he2017neural}, FwFM~\cite{pan2018field}, FiBiNet~\cite{huang2019fibinet}, and  IPNN~\cite{qu2018product}. All experimental results are available on GitHub.

Specifically, a base model $\mathbf{M}$ equipped with FR module $\mathbf{N}$ is represented as $\mathbf{M_N}$ as an augmented model. For example, FM equipped with SENET can be represented as $\mathbf{FM_{SENET}}$. Moreover, for the parallel models with two separate feature refinement modules, as shown in Figure \ref{fig:add_refi}(C), we use $\mathbf{M(2)_{N}}$ to denote it. For example, $\mathbf{DeepFM(2)_{SENET}}$ indicates that DeepFM is equipped with two separate SENET for each sub-network (FM and DNN). In addition, \textbf{SKIP} represents $\mathbf{E} = \mathbf{SKIP}(\mathbf{E})$. For example, $\mathbf{FM_{SKIP}}$ is equal to FM.

\begin{table}[t] 
    \setlength{\abovecaptionskip}{0.2cm}
    \setlength{\belowcaptionskip}{-0.2cm}
\centering
\caption{Statistics of two datasets used in this paper.}
\label{Tab.dataset}
\scalebox{0.77}{
\begin{tabular}{c|c|ccc|cc} 
\hline\hline
Datasets & Positive  & \#Training & \#Validation & \#Testing & \#Fields & \#Features   \\ 
\hline
Criteo  & 26\%  & 36,672,495      &4,584,061   &4,584,061   &39  & 1,086,784  \\
Frappe  & 33\%   &230,889     &28,860    &28,860   &10  & 5,382  \\
\hline\hline
\end{tabular}
}
\end{table}

\begin{table*}
    \setlength{\abovecaptionskip}{0.2cm}
    \setlength{\belowcaptionskip}{-0.2cm}
\centering
\caption{Overall performance comparisons on Criteo and Frappe. The trend of Logloss is similar to AUC, and considering the page limitation, we only present AUC here. We mark the top 5 \textit{Ave.Imp}.}
\label{tab:overall_performance}
\arrayrulecolor{black}
\scalebox{0.80}{
\begin{tabular}{c|cccccccccccccc} 
\hline
\multicolumn{15}{c}{{\cellcolor[rgb]{0.753,0.753,0.753}}\textbf{Criteo}}  \\
\hline
\textbf{Modules} & \textbf{ SKIP } & \textbf{ FEN } & \textbf{ SENET } & \textbf{ FWN } & \textbf{ DFEN } & \textbf{ DRM } & \textbf{ VGate } & \textbf{ BGate } & \textbf{ SelfAtt } & \textbf{ TCE } & \textbf{ PFFN } & \textbf{ GFRL } & \textbf{ FRNet-V } & \textbf{ FRNet-B }  \\ 
\arrayrulecolor{black}\hline
FM           & 0.8080         & 0.8100       & 0.8102       & 0.8100       & 0.8117       & 0.8107       & 0.8090       & 0.8091       & 0.8099       & 0.8112       & 0.8129        & 0.8134        & 0.8139        & 0.8140         \\
DeepFM       & 0.8121         & 0.8128       & 0.8125       & 0.8125       & 0.8121       & 0.8118       & 0.8125       & 0.8127       & 0.8112       & 0.8123       & 0.8129        & 0.8137        & 0.8140        & 0.8141         \\
DeepFM(2)    & 0.8121         & 0.8130       & 0.8128       & 0.8129       & 0.8123       & 0.8119       & 0.8128       & 0.8131       & 0.8129       & 0.8128       & 0.8132        & 0.8138        & 0.8142        & 0.8142         \\
\hline
CN           & 0.8093         & 0.8102       & 0.8095       & 0.8094       & 0.8121       & 0.8109       & 0.8107       & 0.8110       & 0.8102       & 0.8122       & 0.8130        & 0.8139        & 0.8143        & 0.8144         \\
DCN          & 0.8125         & 0.8130       & 0.8116       & 0.8127       & 0.8127       & 0.8118       & 0.8124       & 0.8127       & 0.8122       & 0.8126       & 0.8131        & 0.8142        & 0.8143        & 0.8145         \\
DCN(2)       & 0.8125         & 0.8136       & 0.8122       & 0.8126       & 0.8131       & 0.8120       & 0.8124       & 0.8127       & 0.8129       & 0.8133       & 0.8132        & 0.8144        & 0.8144        & 0.8146         \\
\hline
AFN          & 0.8099         & 0.8140       & 0.8104       & 0.8106       & 0.8116       & 0.8110       & 0.8103       & 0.8101       & 0.8110       & 0.8122       & 0.8130        & 0.8130        & 0.8139        & 0.8141         \\
AFN+         & 0.8108         & 0.8141       & 0.8111       & 0.8124       & 0.8119       & 0.8125       & 0.8119       & 0.8118       & 0.8129       & 0.8128       & 0.8132        & 0.8141        & 0.8141        & 0.8141         \\
AFN+(2)      & 0.8108         & 0.8142       & 0.8116       & 0.8128       & 0.8127       & 0.8129       & 0.8124       & 0.8126       & 0.8131       & 0.8131       & 0.8134        & 0.8142        & 0.8143        & 0.8145         \\
\hline
CN2          & 0.8121         & 0.8119       & 0.8119       & 0.8140       & 0.8131       & 0.8140       & 0.8128       & 0.8133       & 0.8131       & 0.8138       & 0.8130        & 0.8143        & 0.8141        & 0.8143         \\
DCNV2        & 0.8128         & 0.8129       & 0.8122       & 0.8142       & 0.8134       & 0.8141       & 0.8130       & 0.8138       & 0.8135       & 0.8136       & 0.8130        & 0.8143        & 0.8141        & 0.8143         \\
DCNV2(2)     & 0.8128         & 0.8136       & 0.8125       & 0.8140       & 0.8135       & 0.8142       & 0.8139       & 0.8140       & 0.8137       & 0.8138       & 0.8131        & 0.8144        & 0.8143        & 0.8144         \\        
\hline
 \textit{Ave.Imp} & -  &0.18\%\textbf{(5)} &0.03\% &0.13\% &0.15\% &0.12\% &0.09\% &0.11\%&	0.11\% &0.18\%\textbf{(5)} &0.22\%(4) &0.33\%\textbf{(3)} &0.35\%\textbf{(2)} &0.37\%\textbf{(1)} \\
\hline
\multicolumn{15}{c}{{\cellcolor[rgb]{0.753,0.753,0.753}}\textbf{Frappe}}  \\ 
\hline
\textbf{Modules} & \textbf{ SKIP } & \textbf{ FEN } & \textbf{ SENET } & \textbf{ FWN } & \textbf{ DFEN } & \textbf{ DRM } & \textbf{ VGate } & \textbf{ BGate } & \textbf{ SelfAtt } & \textbf{ TCE } & \textbf{ PFFN } & \textbf{ GFRL } & \textbf{ FRNet-V } & \textbf{ FRNet-B }  \\ 
\hline
FM           & 0.9786         & 0.9789       & 0.9800       & 0.9808       & 0.9799       & 0.9820       & 0.9801       & 0.9803       & 0.9806       & 0.9800       & 0.9822        & 0.9821        & 0.9828        & 0.9831         \\
DeepFM       & 0.9824         & 0.9828       & 0.9827       & 0.9824       & 0.9824       & 0.9827       & 0.9828       & 0.9825       & 0.9831       & 0.9824       & 0.9830        & 0.9828        & 0.9837        & 0.9840         \\
DeepFM(2)    & 0.9824         & 0.9830       & 0.9829       & 0.9829       & 0.9827       & 0.9825       & 0.9835       & 0.9828       & 0.9836       & 0.9839       & 0.9829        & 0.9843        & 0.9848        & 0.9846         \\
\hline
CN           & 0.9797         & 0.9829       & 0.9798       & 0.9810       & 0.9810       & 0.9803       & 0.9803       & 0.9803       & 0.9816       & 0.9819       & 0.9826        & 0.9827        & 0.9825        & 0.9826         \\
DCN          & 0.9825         & 0.9830       & 0.9822       & 0.9826       & 0.9838       & 0.9834       & 0.9829       & 0.9820       & 0.9829       & 0.9827       & 0.9828        & 0.9838        & 0.9838        & 0.9837         \\
DCN(2)       & 0.9825         & 0.9834       & 0.9829       & 0.9831       & 0.9843       & 0.9843       & 0.9835       & 0.9829       & 0.9832       & 0.9839       & 0.9838        & 0.9840        & 0.9844        & 0.9847         \\
\hline
AFN          & 0.9812         & 0.9826       & 0.9812       & 0.9816       & 0.9822       & 0.9821       & 0.9821       & 0.9814       & 0.9820       & 0.9826       & 0.9815        & 0.9835        & 0.9838        & 0.9838         \\
AFN+         & 0.9827         & 0.9838       & 0.9827       & 0.9831       & 0.9840       & 0.9836       & 0.9830       & 0.9826       & 0.9830       & 0.9836       & 0.9827        & 0.9838        & 0.9843        & 0.9844         \\
AFN+(2)      & 0.9827         & 0.9840       & 0.9830       & 0.9840       & 0.9846       & 0.9838       & 0.9839       & 0.9827       & 0.9837       & 0.9838       & 0.9834        & 0.9841        & 0.9844        & 0.9847         \\
\hline
CN2          & 0.9810         & 0.9822       & 0.9813       & 0.9826       & 0.9830       & 0.9825       & 0.9827       & 0.9813       & 0.9827       & 0.9821       & 0.9817        & 0.9825        & 0.9826        & 0.9834         \\
DCNV2        & 0.9830         & 0.9833       & 0.9835       & 0.9831       & 0.9839       & 0.9837       & 0.9833       & 0.9826       & 0.9829       & 0.9833       & 0.9831        & 0.9840        & 0.9839        & 0.9845         \\
DCNV2(2)     & 0.9830         & 0.9838       & 0.9838       & 0.9838       & 0.9844       & 0.9838       & 0.9837       & 0.9828       & 0.9832       & 0.9841       & 0.9835        & 0.9845        & 0.9841        & 0.9849         \\    

\hline
\textit{Ave.Imp} & -& 0.10\%& 0.04\%& 0.08\%& 0.12\%\textbf{(4)}& 0.11\%\textbf{(5)}         & 0.08\%           & 0.02\%           & 0.09\%             & 0.11\%         & 0.10\%          & 0.17\%\textbf{(3)}          & 0.20\%\textbf{(2)}             & 0.22\%\textbf{(1)}  \\       
\hline
\hline
\end{tabular}}
\end{table*}

\textbf{Reproducibility:}
We implement all the above-mentioned models with Pytorch~\cite{paszke2019pytorch}. All models are learned by optimizing the Cross-Entropy loss with Adam~\cite{kingma2014adam} optimizer. We implement the Reduce-LR-On-Plateau scheduler during the training process to reduce the learning rate by a factor of 10 when the specified metric stops improving for four consecutive epochs. The learning rate is fine-tuned in \{0.1, 0.01, 0.001\}. Early stopping is used to avoid overfitting when the AUC on the validation set stops improving for 3 consecutive epochs. The mini-batch size is searched in \{2000, 5000, 10000\}. The default embedding size is 16 for Criteo and Frappe. Following previous works~\cite{huang2019fibinet,cheng2020adaptive, guo2017deepfm, song2019autoint}, we set the same structure (i.e., 3-layers MLP, 400-400-400) for the models that involve DNN for a fair comparison. All activation functions are ReLU unless otherwise specified, and the dropout rate is 0.5. Notably, our aim is to evaluate the effectiveness of different FR modules rather than base models. Therefore, we keep the settings of the base model unchanged for fairness and then adjust the parameters of the FR modules to achieve the best results.

\textbf{Significance Test:} We repeat each augmented model at least 10 times and report the average performance in testing dataset. We also perform a two-tailed pairwise t-test~\cite{bhattacharya2002ttest} to determine the significance of differences between the base models and their augmented versions with different FR modules. All results are \textbf{statistically significant with p-value<0.01}. 

\subsection{Overall Comparison}

\subsubsection{Performance Comparisons} 
We evaluate the performance of different FR modules by integrating them to base models, and we calculate their average improvement (\textit{Ave.Imp}) compared with those base models. A model with $\textbf{SKIP}$ represents the base model only without any augmented FR module. Table \ref{tab:overall_performance} shows the evaluation results on two datasets, and we have the following observations: 

First, for FI-based CTR prediction models, complex feature interaction structures (e.g., CN, AFN, and CN2) can significantly improve the performance of CTR prediction models than low-order interactions (FM). Furthermore, parallel models (e.g., DeepFM, DCNV2) integrating implicit feature interactions (e.g., DNN) can also boost the performance of CTR prediction models, since the parallel models incorporate two or more kinds of interaction information. 

Second, applying FR modules to generate refined feature representations can effectively improve the performance of base CTR models. The performance of 156 augmented models is shown in Table~\ref{tab:overall_performance}, and most of these models significantly outperform corresponding base models on both datasets. Specifically, 142 augmented models achieve better performance, and the average AUC improvement is 0.0021 on Criteo. In addition, the average AUC improvement of 148 augmented models is 0.0012 on Frappe. Furthermore, the $Ave.Imp$ are positive for all 13 modules on both datasets. The above results verify the effectiveness and compatibility of the FR modules. 

Third, refining feature representation is as important as modeling feature interactions. DCNV2 achieves the best performance among all the base models (i.e., 0.8129 and 0.9830 for Criteo and Frappe). However, after using FR modules to refine feature representations, the best-performing augmented models surpass DCNV2, including ones with some simple stacked base models, e.g., $FM_{FRNet-B}$, $CN_{FRNet-B}$, etc. This is because a fixed representation learned by a plain embedding layer is insufficient and limits the FI and prediction layers to achieve better results. Table \ref{tab:overall_performance} verifies that effective feature refinement can greatly boost the base model performance even with sub-optimal FI modules on top. Hence, refining feature representation is a promising direction in the CTR prediction task. We also observe that in most cases, even using FR modules, the performance of augmented models still depends on the FI module, where high-order FI modules outperform low-order modules and parallel models outperform stacked models. This indicates that designing novel FI structures is also necessary. 

Fourth, the performance of different FR modules varies greatly. Although all $Ave.Imp$ are positive, different FR modules have varying effects on CTR models. For instance, GFRL, FRNet-V, and FRNet-B perform more consistently and effectively than other FR modules. They rank in the top three on both datasets and can improve the performance of all base models. These modules share the same characteristics: they use Sigmoid functions for the weights of features and generate new complementary feature representations to compensate for information that is not selected. On the other hand, SENET, VGate, and BGate perform relatively poorly in both Criteo and Frappe, with other FR modules falling in between.  

\begin{figure}
    \setlength{\abovecaptionskip}{0.2cm}
    \setlength{\belowcaptionskip}{-0.2cm}
    \centering
    \includegraphics[width=0.48\textwidth]{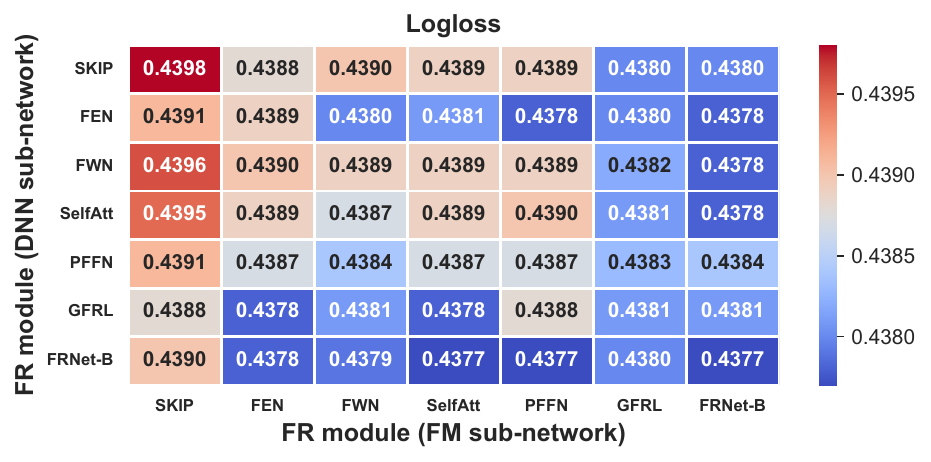}
    \caption{Logloss of DeepFM with two separate FR modules. }
    \label{fig:deepfm_para}
\end{figure}

\subsubsection{Performance of Assigning Two Separate FR Modules}
\label{exp:newdesign}
First, using two separate but the same FR modules for two sub-networks in parallel CTR prediction models performs better than utilizing a shared FR module. By applying 13 FR modules (except SKIP) onto 4 parallel base models (DeepFM, DCN, AFN+, and DCNV2), we have 52 groups of comparisons. 50 of the 52 comparison groups show better performance by assigning two separate FR modules on Criteo, with only 2 group experiments not performing better (i.e., $DCN(2)_{FWN}$ vs. $DCN_{FWN}$ and $DCNV2(2)_{FWN}$ vs. $DCNV2_{FWN}$). On Frappe, 50 groups achieve better performance. Intuitively, separate FR modules can adjust corresponding feature representations adaptively to the needs of each sub-network and thus leading to better performance than using a shared FR module. 


Second, we evaluate the effectiveness of using two different separate FR modules. As shown in Figure~\ref{fig:deepfm_para}, based on DeepFM, we apply different FR modules for two sub-networks, i.e., FM and DNN. For example, <FEN, FRNet-B> represents that FM is applied with FEN and DNN is applied with FRNet-B. We have the following observations: (1) Each sub-network can benefit from the refined features individually and improve the performance of CTR prediction models. Specifically, for <SKIP, $\boldsymbol{\Phi}$> and <$\boldsymbol{\Phi}$, SKIP> ($\boldsymbol{\Phi} \in $ \{FEN, FWN, SelfAtt, PFFN, GFRL, FRNet-B\}), all variants outperform <SKIP, SKIP>. (2) In most cases, better performance is achieved while using FR modules for both sub-networks, e.g., <$\boldsymbol{\Phi}$, FEN> is better than <SKIP, FEN> and <FEN, $\boldsymbol{\Phi}$> is better than <FEN, SKIP>, which can also be observed by replacing FEN with the other FR modules. (3) Many combinations of FR modules can achieve the best performance (0.4377 Logloss), offering the possibility of improving base model performance with a proper combination of FR modules. Meanwhile, we can also choose the suitable combination of modules according to specific needs, e.g., interpretability, parameters or training time.
\begin{table}
\centering
\caption{A summary of model complexities (ignoring the bias term) and training time (per epoch) on Criteo. The hidden layer depth $L_d$ and dimension $m$ used in different modules are different. The hyper-parameters are mainly based on the original paper. We also add two SOTA models (i.e., DeepFM and DCNV2) as comparisons}
 %
\label{tab:complexity}
\scalebox{0.78}{
\begin{tabular}{c|cl|r} 
 \hline
Module or Model & Space Complexity & Parameters &Times(s)   \\
 \hline  
SKIP(Basic FM)&    - &   18,475,329   &  427($\pm$5) \\
 \hline 
FEN     & $\makecell[c]{Dm(F+1)+(m^2+m)L_d}$   & +303,221 &+49($\pm$6)  \\
 \hline 
SENET     & $2F^2$      &+3,120   &+48($\pm$6) \\
 \hline 
FWN& $FD^2+FD$   &+10,608   &  +60($\pm$9)\\
 \hline 
DFEN    &$\makecell[c]{5hDD_k+Dm(F+1)+(m^2+m)L_d}$ &+353,952 &+117($\pm$7)    \\
 \hline
DRM& $3F^2$      &+4,563&  +65($\pm$5)\\
 \hline 
VGate     & $FD$      & +624  &+48($\pm$3) \\
 \hline 
EGate     & $FD^2$    &+9,984    & +55($\pm$4)  \\
 \hline 
TCE& $2FDm$ & +39,936 &    +62($\pm$5)   \\
 \hline 
 PFFN&$2FDm+2L_dFD^2$ & +41,472   &+104($\pm$5)   \\
 \hline 
SelfAtt   &$4hDD_k$     &+17,152    &+106($\pm$6)  \\
 \hline 
 GFRL   &$2(FD(D+1)+Dm(F+1))$     &+185,344   &+56($\pm$9)   \\
 \hline 
FRNet-V&$\makecell[c]{2(4D^2+Dm(F+1)+L_d(m^2+m))}$&+166,818 &+68($\pm$6) \\
 \hline 
FRNet-B& $\makecell[c]{2(4D^2+Dm(F+1)+L_d(m^2+m))}$ &+166,818&+72($\pm$4) \\
\hline
\hline
DeepFM      & - &+573,601&+66($\pm$4)   \\ 
\hline
DCNV2      & - &+657,440 &+86($\pm$9)      \\
%
\bottomrule
\end{tabular}
}
\end{table}

\begin{figure}
    \setlength{\abovecaptionskip}{0.2cm}
    \setlength{\belowcaptionskip}{-0.2cm}
    \centering
    \includegraphics[width=0.42\textwidth]{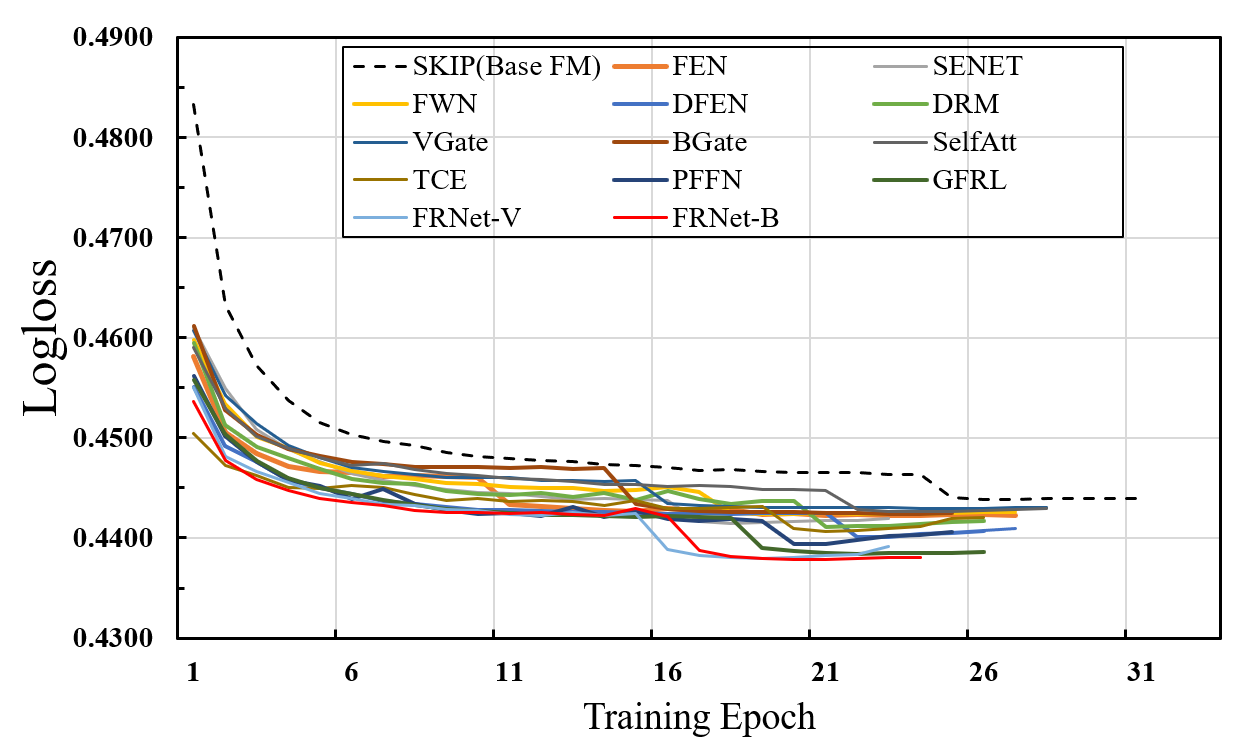}
    \caption{Training speed of different modules. }
    \label{fig:training speed}
\end{figure}

\begin{figure*}[tb]
    \setlength{\abovecaptionskip}{0.2cm}
    \setlength{\belowcaptionskip}{-0.2cm}
\centering
\subfloat[BGate (VGate, FWN)]{
\begin{minipage}[t]{0.20\linewidth}
\centering
\includegraphics[width=1.0\textwidth]{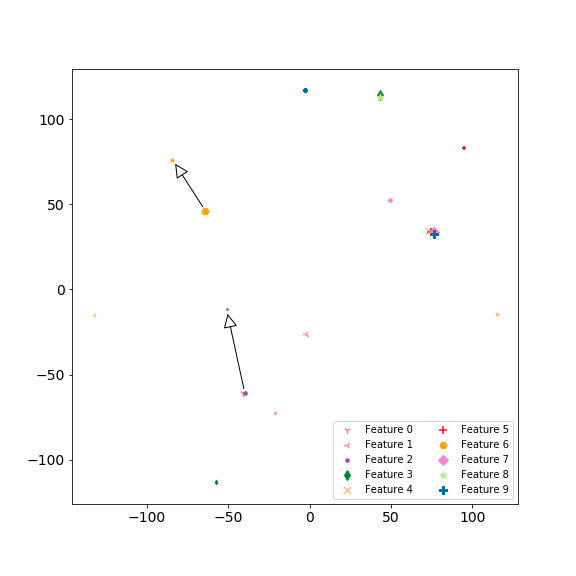}
\end{minipage}
}
\subfloat[FEN (DFEN, SENET)]{
\begin{minipage}[t]{0.20\linewidth}
\centering
\includegraphics[width=1.0\textwidth]{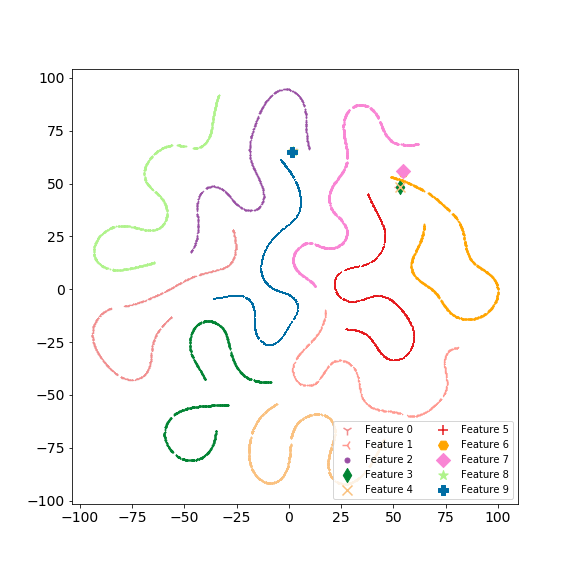}
\end{minipage}
}
\subfloat[TCE]{
\begin{minipage}[t]{0.20\linewidth}
\centering
\includegraphics[width=1.0\textwidth]{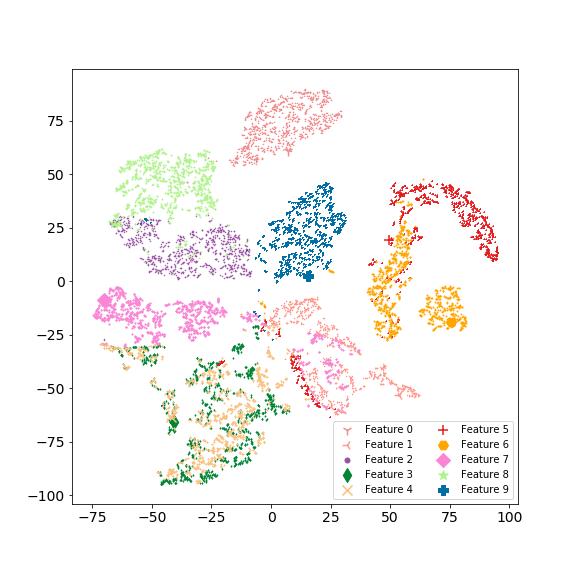}
\end{minipage}
}
\subfloat[SelfAtt (PFFN, DRM)]{
\begin{minipage}[t]{0.20\linewidth}
\centering
\includegraphics[width=1.0\textwidth]{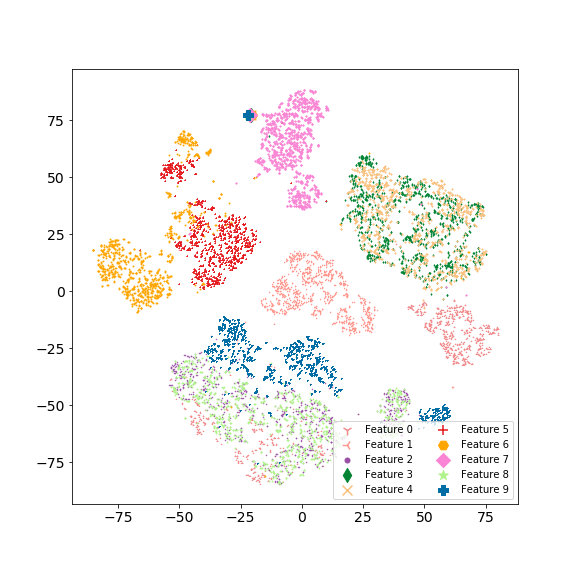}
\end{minipage}
}
\subfloat[FRNet-B (GFRL,FRNet-V)]{
\begin{minipage}[t]{0.20\linewidth}
\centering
\includegraphics[width=1.0\textwidth]{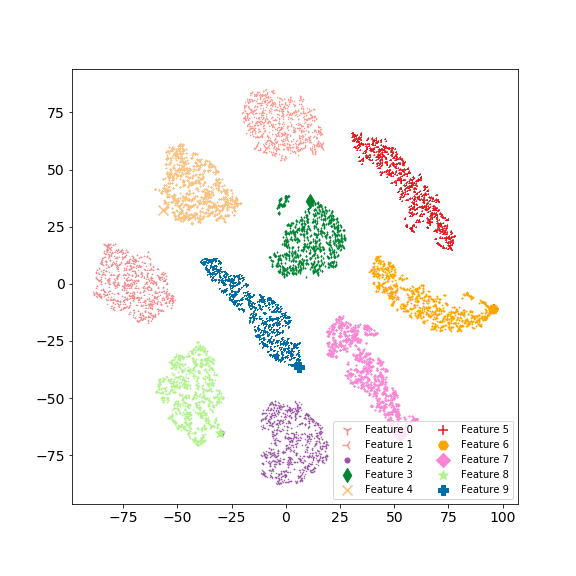}
\end{minipage}
}
\caption{Visualization of five typical context-aware feature representations. }
\label{fig:vis}
\end{figure*}

\begin{figure*}[tb]
    \setlength{\abovecaptionskip}{0.2cm}
    \setlength{\belowcaptionskip}{-0.2cm}
\centering
\subfloat[FWN, Field 23 and 30 (Before)]{
\begin{minipage}[t]{0.24\linewidth}
\centering
\includegraphics[width=0.9\textwidth]{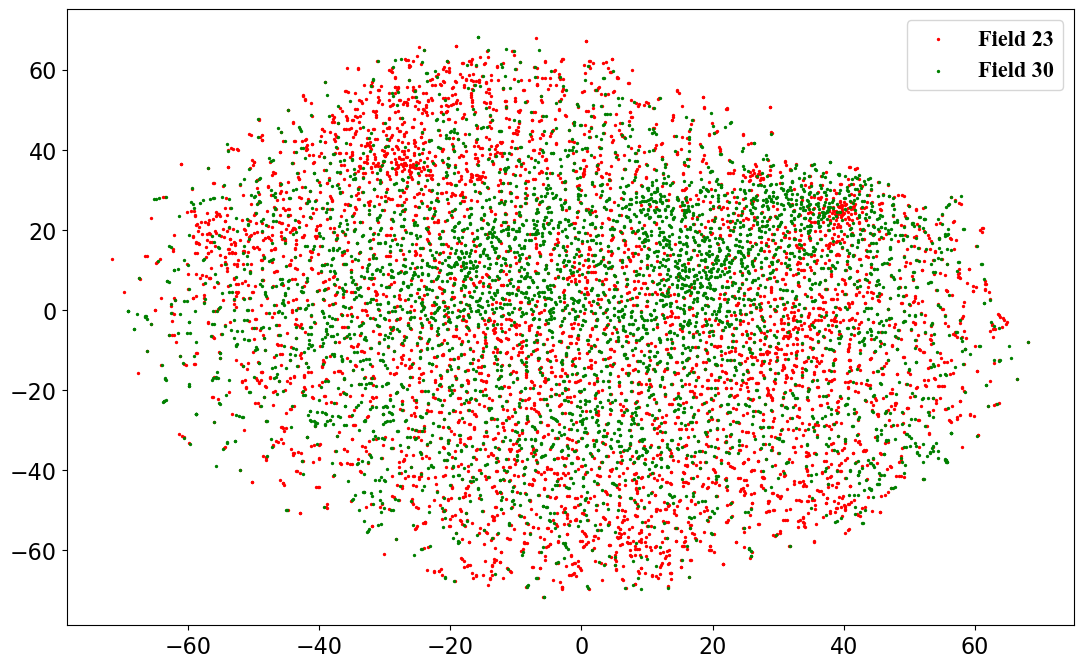}
\end{minipage}
}
\subfloat[FWN, Field 23 and 30 (After)]{
\begin{minipage}[t]{0.24\linewidth}
\centering
\includegraphics[width=0.9\textwidth]{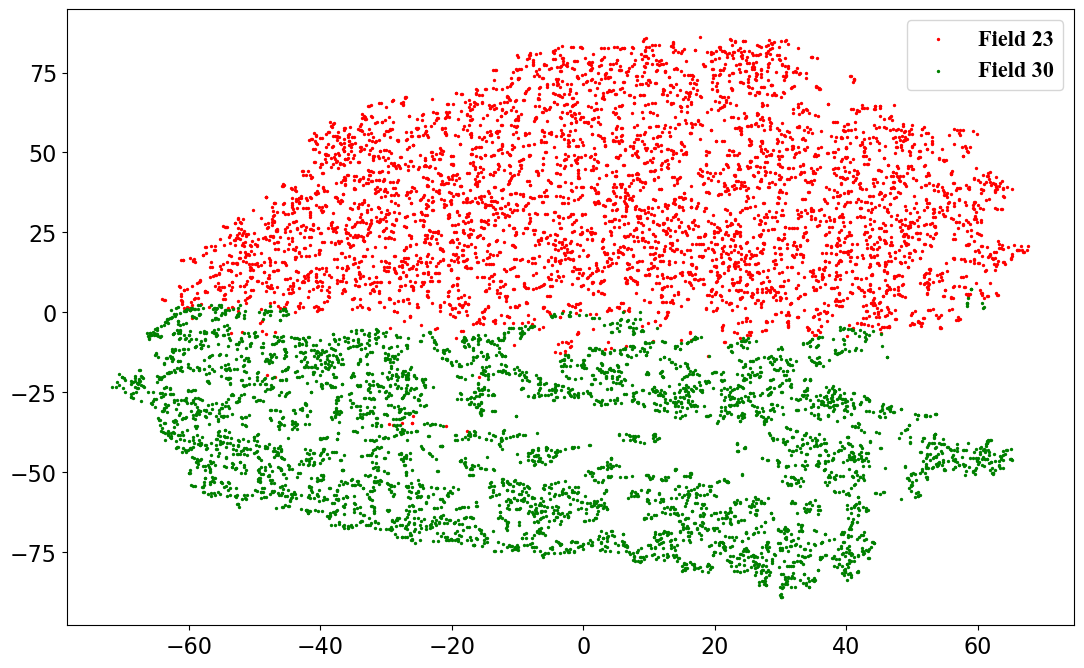}
\end{minipage}
}
\subfloat[VGate, Field 13 and 31 (Before)]{
\begin{minipage}[t]{0.24\linewidth}
\centering
\includegraphics[width=0.9\textwidth]{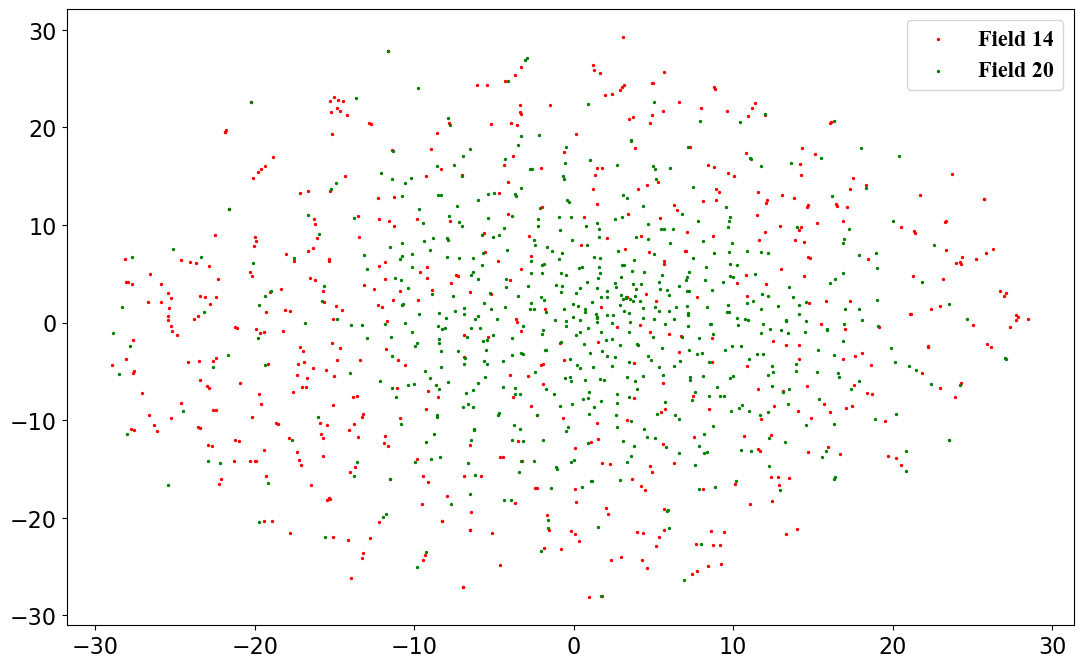}
\end{minipage}
}
\subfloat[VGate, Field 13 and 31 (After)]{
\begin{minipage}[t]{0.24\linewidth}
\centering
\includegraphics[width=0.9\textwidth]{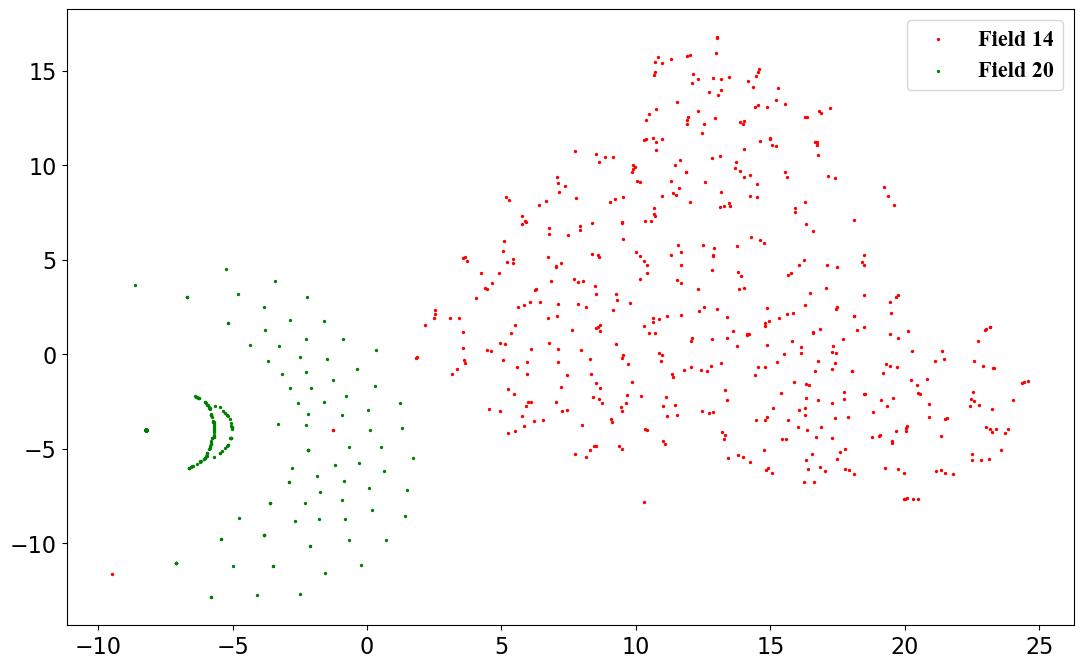}
\end{minipage}
}

\centering
\caption{Visualization of different fields' embeddings before and after the processing of FWN and VGate.}
\label{fig:trans}
\end{figure*}

\subsection{Training Efficiency}
\subsubsection{Complexity Analysis.} Using FM as the most lightweight base model, we measure each FR module's added parameters and training time. We also estimate the theoretical space complexity, shown in Table \ref{tab:complexity}. Generally, FR modules do not significantly increase the number of parameters compared to feature embeddings in the base FM model. Better-performing FR modules, such as DFEN, GFRL, FRNet-V, and FRNet-B, tend to require more parameters. Meanwhile, applying FR modules increases the training time, with the average increase ranging from 49 to 117 seconds. However, augmented models (e.g., $FM_{GFRL}$, $FM_{FRNet-B}$) using FR modules outperform SOTA FI-based models (e.g., DeepFM, DCNV2) with fewer parameters and similar training time, demonstrating the effectiveness of FR modules.

\subsubsection{Convergence Speed}
Compared to the basic FM model, adding FR modules accelerates the convergence speed. Figure \ref{fig:training speed} plots the test Logloss of these augmented models based on different FR modules during the training process. Firstly, adding FR modules can accelerate the convergence of the basic FM model. Thus, although adding FR modules increases the training time per epoch, the overall training time is similar because those augmented models converge faster with fewer training epochs. Secondly, Second, adding the FR module allows for faster and better performance than the base FM model. Specifically, within ten epochs, the performance of most augmented models outperforms basic FM models. The possibility is that the weight learning module in most FR modules essentially approximates or contains the gating components, which are responsible for selecting useful embedding information from the original embedding and feeding them to the FI layer. As mentioned in GateNet~\cite{huang2020gatenet}, GDCN~\cite{wang2023gdcn} and EDCN~\cite{chen2021enhancing}, adopting gating mechanisms can accelerate the convergence of the model.

\subsection{Visualization of Feature Representations}
In this section, we visualize the refined feature representations with t-SNE~\cite{van2008visualizing}. We select 10 features from the same field and sample 1,000 instances for each feature from Criteo. As shown in Figure \ref{fig:vis}, different color represents the 10 features and their 1,000 refined representations. The larger symbols (e.g., dots, squares, etc.) show the original feature representations, while the smaller symbols denote 1,000 different context-aware feature representations. Based on the properties of context-aware and non-linearity, we summarize the following patterns: 

1) FR modules, such as FWN, VGate, and BGate, cannot learn context-aware refined representations, but they transform the original feature representations into another high-dimensional space. Figure \ref{fig:vis}(a) shows this with two arrows. Although these modules have fewer parameters than other FR modules, they are less effective in comparison. Furthermore, further visualization analysis reveals that these methods can make the embeddings within different fields more discriminative by modeling the intra-field information, which is consistent with the findings in NON~\cite{zhao2021non}. As shown in Figure \ref{fig:trans}, we plot all feature representations belonging to two selected fields by t-SNE, and Figure \ref{fig:trans}(a) and (b) represent before and after processing, respectively. Obviously, after the processing of FWN and VGate(i.e., Figure \ref{fig:trans} (b) and (d)), the features within the same fields are closer to each other, and the features within different fields are easier to distinguish.

2) FR modules, such as FEN, DFEN, and SENET, are able to learn context-aware but linearity refined representations. These modules adopt the selection paradigm and learn vector-level weights for features in different instances. Their refined representations have a strictly linear relationship with the corresponding original feature representations in high-dimensional space, as depicted by a curve in low-dimensional space in Figure \ref{fig:vis}(b). They generally perform less effectively as well due to the limitation of linearity. However, vector-lever weight learning is beneficial for interpretation, as the weights directly show the importance of a feature in different instances.

\begin{figure*}[tb]
\centering
\subfloat[FEN]{
\begin{minipage}[t]{0.25\linewidth}
\centering
\includegraphics[width=0.9\textwidth]{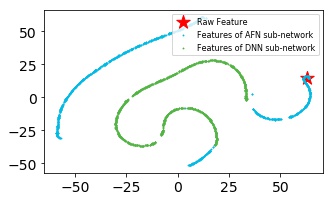}
\end{minipage}
}
\subfloat[SENET]{
\begin{minipage}[t]{0.25\linewidth}
\centering
\includegraphics[width=0.9\textwidth]{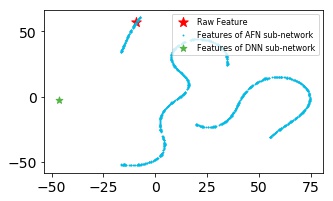}
\end{minipage}
}
\subfloat[TCE]{
\begin{minipage}[t]{0.25\linewidth}
\centering
\includegraphics[width=0.9\textwidth]{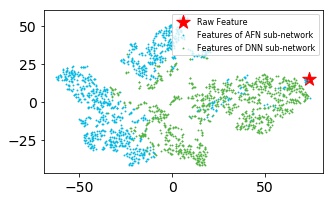}
\end{minipage}
}
\subfloat[FRNet-B]{
\begin{minipage}[t]{0.25\linewidth}
\centering
\includegraphics[width=0.9\textwidth]{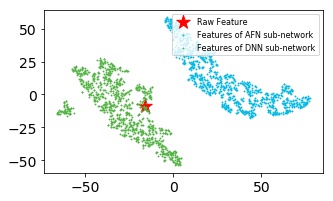}
\end{minipage}
}

\centering
\caption{Visualization of two groups refined representations generated by two separate FR modules applied before two sub-networks in AFN+(AFN and DNN). Red pentagram is the sampled feature, and other representations are generated based on it.}
\label{fig:weight}
\end{figure*}

3) FR modules, such as DRM, SelfAtt, TCE, PFFN, GFRL, FRNet-V, and FRNet-B, learn context-aware and non-linearity refined representations, as shown in  Figure \ref{fig:vis}(c)-(e). These FR modules differ in terms of activation functions and the generation paradigm. (i) TCE learns bit-level weights with the linear activation function, which increases the degrees of freedom and overcomes the linearity limitation compared to vector-level weight learning. (ii) SelfAtt, PFFN, and DRM use transformers to generate context-aware feature representations and avoid linearity issues. However, in Figure \ref{fig:vis}(d), the feature representations generated by these FR modules still heavily overlap, i.e., different color clusters are overlapping, showing room for improvement (iii) For FRNet-B, FRNet-V, and GFRL, their context-aware nonlinear representations have clear and distinct feature spaces, with clusters of the same color not overlapping. They use Sigmoid activation to restrict the range of weights from 0 to 1, ensuring the refined representations are always close to the original representation. Therefore, these three modules significantly outperform other FR modules by having more distinct refined representations of different features.


\subsection{Feature Distributions for Parallel Models}
To examine the discriminative feature distributions for the two FR modules within parallel models, we select 4 representative FR modules based on AFN+. Specifically, AFN+ consists of AFN and DNN, where AFN uses a logarithmic transformation layer to learn arbitrary-order cross features, and DNN captures high-order element-wise interactions. As shown in Figure \ref{fig:weight}, we choose the same feature (red pentagram), and 500 instances containing the selected feature. We can generate two groups of 500 corresponding context-aware representations (blue and green clusters), representing the refined feature representations generated by the two FR modules inserted before AFN and DNN. We have the following observations:

First, each sub-network of the parallel models needs different feature representation distribution. The refined feature representations adjusted by two FR modules form two separate clusters (represented by blue and green dots). We conduct similar experiments for other features, which show similar results. Table \ref{tab:overall_performance} and Section \ref{sec:experiment} have verified that having distinct feature representations achieves better performance than having a single representation, highlighting the reasonability of assigning separate FR modules rather than only using a shared module.

Second, different FR modules show different representation characteristics. FEN and SENET only produce context-aware but linearity-refined feature representations for the two groups of features. In Figure \ref{fig:weight}(b), the refined feature representations generated by DNN only have an identical feature. Upon examining the weights learned to generate those refined representations, we find that all the weights are 0, which indicates that SENET considers the selected feature unimportant for the DNN sub-networks. In addition, SENET has the worst performance compared with other FR modules. On the other hand, for TCE and FRNet-B, the refined feature representations (green cluster) of the DNN sub-network are closer to the original feature representation (red pentagram) and clearly distinguishable, showing the quality of separate feature representations.


\section{Future Directions}
We present four potential research directions for further research in feature refinement and CTR prediction.

\textbf{Auto-tuning of Hyper-parameters.} Finding the optimal hyper-parameters for deep learning-based models is a critical issue, especially for CTR prediction models where the best hyper-parameters can differ depending on the dataset. The integration of FR modules into CTR prediction models increases the number of hyper-parameters, making the search process more complex and time-consuming using traditional grid search methods. Therefore, advanced AutoML techniques that are suitable for CTR prediction with FR integrated should be developed to streamline the search process and enhance the performance of these models.

\textbf{Neural Architecture Search (NAS).} Designing an effective FR module depends on expert knowledge and experience. A pre-determined network structure may not be optimal for different base models and datasets. Recent advancements in the NAS area for automating the design of FI structures and CTR models~\cite{liu2020autogroup, meng2021autopi} have shown promising results. Therefore, using NAS technology to design FR modules is a feasible direction, given the lack of existing work in this area. Further research into advanced NAS techniques is expected to improve the design process of FR modules.

\textbf{User Behavior Modeling.} The modeling of user behavior is an important aspect of CTR prediction in recent studies. The history of user behavior holds valuable information about a user's interests and significantly influences predicting the probability of clicks on future items. However, current FR modules have not been effectively applied to CTR models that model user behavior~\cite{zhou2018din,zhou2019dien,pi2020search_sim,ren2019lifelong}. The behavioral features are typically sequential and ordered by time, so designing appropriate FR modules for these models is a significant challenge.

\textbf{Evaluation or Interpretation Methods}. Evaluating and explaining the impact of different FR modules remains a major challenge. Currently, the effectiveness of a module is mainly evaluated by the improvement in performance after it is added to a base model. Although some studies, such as~\cite{huang2020gatenet, wang2022frnet}, claim that the FR layer can lead to improved feature representations, there is still a lack of clear qualitative or quantitative metrics to determine what makes the refined feature representations effective.

\section{Conclusions}
In recent years, refining feature representations have emerged as a promising area to enhance the performance of existing CTR models. In this paper, we provide the first comprehensive and systematic evaluation of existing FR modules using both qualitative and quantitative methods. Our study bridges the need for more research between the feature interaction layer and the feature embedding layer. Our experiments demonstrate the significant impact of feature refinement for the basic CTR models, and further research in this area is highly encouraged. Meanwhile, we present a new architecture of assigning independent FR modules for parallel models and verify its rationality and effectiveness. We hope that our summarization and evaluation provide valuable insights for researchers to advance the field of CTR prediction further and beyond.


\clearpage
\newpage
\bibliographystyle{ACM-Reference-Format}
\bibliography{survey1}
\end{document}